\documentstyle[12pt]{article}
\input{epsf}

\newcommand{\beq}{\begin{eqnarray}}
\newcommand{\eeq}{\end{eqnarray}}

\newcommand{\ket}[1]{| #1 \rangle}
\newcommand{\bra}[1]{\langle #1 |}

\newcommand{\pslash}{{p\hspace{-5pt}/}}
\newcommand{\qslash}{{q\hspace{-5pt}/}}

\newcommand{\gpNN}{g_{\pi NN}}
\newcommand{\geNN}{g_{\eta NN}}
\newcommand{\gpNR}{g_{\pi NN^*}}
\newcommand{\geNR}{g_{\eta NN^*}}

\def\lsim{\displaystyle\mathop{<}_{\sim}}

\begin{document}

\begin{center}
{\large \bf Suppression of $\pi NN(1535)$ Coupling
in the QCD Sum Rule}\footnote{Talk given by A.H. at YITP workshop
''Recent 
developments in QCD and hadron physics", Kyoto, December (1996)}\\
\vspace*{0.5cm}
{\it A. Hosaka\footnote{e-mail address: hosaka@la.numazu-ct.ac.jp}  \\
Numazu College of Technology \\
3600 Ooka, Numazu 410 Japan \\
and\\
D. Jido and M. Oka \\
Department of Physics, Tokyo Institute of Technology\\
Meguro, Tokyo 152  Japan}
\end{center}

\vspace*{0.5cm}
\begin{abstract}
The $\pi NN^*$ and $\eta NN^*$ coupling constants are studied
in the QCD sum rule ($N^* \equiv N(1535)$).
We investigate the two point functions between the
vacuum and a one meson state in the soft meson limit.
The operator product expansion is explicitly 
performed up to terms of dimension eight.
We find that $\gpNR$ vanishes to this order, while
$\geNR$ has a finite contribution.  
The vanishing of $\gpNR$ is a rigorous consequence in the chiral limit. 
The results of the QCD sum rule are compared with those of low energy
effective models.
\end{abstract}

There is a considerable interest in baryon resonances to test
effective models for hadrons.  
Much of the motivations are stimulated by the fact that 
detailed measurements of 
resonance properties will soon become available at facilities such 
as TJNAF (former CEBAF).  
In particular, 
expecting data on form factors of electromagnetic and strong 
interactions will be extremely useful in understanding details of hadron 
structure.  

Among various baryon resonances, the negative parity baryons,
$N(1535)$, $\Lambda(1670)$ and $\Sigma(1750)$
(in the following
we denote those resonances by $B^{*}$), possess interesting
properties.
The most distinguished feature is 
their relatively large decay modes of $B^* \to \eta B$.
Because of the smallness of the available phase space, 
this fact suggests that relatively
large coupling constants of $\eta BB^*$ as compared to those of 
$\pi BB^*$~\cite{PDG}.  

One may also look at the problem in the following way.  
Using the experimental decay widths of the resonances, we obtain, for 
example, 
$\gpNR \sim 0.7$ and $\geNR \sim 2$.  
These values are in fact much smaller than those in the $NN$ sector:  
$\gpNN \sim 13$ and $\geNN \leq 5$.  
Furthermore, the pion couples weaker than the eta in the $NN^*$ sector,
as 
opposed to the $NN$ sector.  
Thus, one may ask why the coupling $\gpNR$ is suppressed so much 
as compared with other couplings.  

Recently, Jido, Kodama and Oka
studied the masses of the octet
baryons of negative parity in the QCD sum
rule~\cite{jko}.  
An interesting observation is that mass differences between $B$ and
$B^*$ 
are related to the strength of chiral symmetry breaking 
($\sim \langle q \bar q \rangle$), suggesting that $B$ and $B^*$ are 
chiral partners of spontaneously broken chiral symmetry.  
Then it would be very interesting to see if the above relations 
among the coupling constants are derived from chiral symmetry or not.

Following Shiomi and Hatsuda~\cite{sh}, 
let us study the two point function between the vacuum and a one meson
state in 
the soft meson limit $(q^{\mu} \to 0)$:
\begin{equation}
   \Pi^{m^\alpha}(p) = i \int d^4 x \, e^{ip\cdot x}
        \bra{0} TJ_{N}(x;t) \bar{J_{N}}(0;t^\prime) \ket{m^\alpha (q \to
0)}  
   \label{eq:cor}  \nonumber \\
\end{equation}
where the nucleon interpolating field $J_{N}$ is defined by
\begin{eqnarray}
   J_{N}(x;t) & = & \varepsilon^{abc} [(u_{a}(x)Cd_{b}(x))
   \gamma_{5} u_{c}(x) + t (u_{a}(x) C \gamma_{5} d_{b}(x))
   u_{c}(x)] \, ,
        \label{eq:nucur}
\end{eqnarray}
and $m^\alpha$ denotes either $\pi$ ($\alpha = 1,2,3$) or $\eta$ 
($\alpha = 0$).  
When $t = -1$ the current (\ref{eq:nucur}) reduces to the Ioffe current
which 
is supposed to couple strongly to the ground state nucleon~\cite{i}.  
As discussed in ref.~\cite{jko}, a large coupling strength of the
current 
with negative parity states is achieved when $t = 0.8$.  
Therefore, for transition matrix elements between $N$ and $N^*$, 
one of the currents in (\ref{eq:cor}) is chosen to be the Ioffe current 
with $t=-1$ and the other one with $t=0.8$.

The soft meson limit might not be a good approximation for 
matrix elements between $N$ and $N^*$, and also for the coupling of eta.
However, it would be always instructive to see how things look like in 
this limit.  

Let us first look at the phenomenological expression of the correlation
function to see how information of the negative parity nucleon
can be extracted.
We shall see this for the $\pi NN^*$ coupling.
Using the phenomenological $\pi NN^*$
interaction lagrangian
$ 
{\cal L}_{\pi NN^*} = \gpNR \bar{N}^{*} \tau_{a} \pi^{a} N,
$
with the standard notations, 
$\pi NN^{*}$ contributions in the
$\Pi^{\pi^\alpha}(p)$ is given in the soft pion limit by
\begin{equation}
   \label{eq:pNN*}
    \gpNR \lambda_{N} \lambda_{N^{*}}
    \left[{p^{2} + m_{N}m_{N^{*}} \over (p^{2} - m_{N}^{2}) (p^{2} -
    m_{N^{*}}^2)} + { \pslash (m_{N} + m_{N^{*}}) \over (p^{2} -
    m_{N}^{2}) (p^{2} - m_{N^{*}}^2)}\right]i\gamma_{5} \tau_\alpha \, , 
\end{equation}
where $\lambda_{N}$ and $\lambda_{N^{*}}$ are defined by
$\langle 0 | J_{N} | N\rangle = \lambda_{N} u_{N}$ and 
$\bra{0} J_{N} \ket{N^{*}} = i \lambda_{N^*} \gamma_5 u_{N^*}$, 
respectively, with
$u_{N}$ and $u_{N^*}$ being the Dirac spinor for $N$ and $N^*$.  
We note that there are two terms in (\ref{eq:pNN*}), which is contrasted 
with the $\pi NN$ case, where there is only one term:
\begin{equation}
   \gpNN \lambda_{N}^{2} {i \gamma_{5} \tau_\alpha \over
    p^{2} - m_{N}^{2} } \, .    \label{eq:pheNN}
\end{equation}
In fact, this result is obtained from (\ref{eq:pNN*}) by replacing 
$m_{N^*}$ by $-m_{N}$ there.  
The minus sign appears here because the ordering of $\gamma_5$ and 
$\pslash$ is reversed in the expressions for $\gpNN$ and $\gpNR$.

In the soft pion limit,
Shiomi and Hatsuda studied the sum rule for the non-vanishing $\gamma_5$ 
term of (\ref{eq:pheNN})~\cite{sh}.
Recently, Birse and Krippa studied the coupling constant
$\gpNN$ at a nonzero pion momentum $q$ by looking at a term 
proportional to $\qslash$~\cite{bk}.
In our case for the $\pi NN^*$ coupling constant, 
we shall study the term proportional to $\pslash \gamma_{5}$ in the soft 
meson limit,
which is expected to have
a dominant contribution from the $\pi NN^*$ coupling.
In fact, there could be contributions from positive parity resonances,
and 
a dominant part would be from the lowest resonance $N(1440)$.
Such a term is, however, proportional to
the mass difference $M_{N(1440)} - M_{N}$ unlike the sum as in
the second term of (\ref{eq:pNN*}).
Thus the contribution from $N(1440)$ will be relatively
suppressed as compared with that of $N(1535)$.
Moreover since we choose the
mixing parameter $t \sim 0.8$ such that
the interpolating field (\ref{eq:nucur})
couples strongly with negative parity states, we expect least
contamination from positive parity resonances.
Similarly, the sum rule for the $\eta NN^{*}$ coupling is
constructed
by replacing the isospin matrices $\tau^\alpha$ in the $\pi NN^*$
coupling by the unit matrix.

The correlation function is now computed by the operator product
expansion (OPE) perturbatively in the deep Euclidean region.
The $i \pslash \gamma_{5}$ term takes the following form
\begin{eqnarray}
    \Pi^{\rm OPE}(p) & = & i \int d^{4}x \, e^{ip\cdot x} \, \bra{0}{\rm
T}
      J_{N}(x;s) \bar{J}_{N}(0;t) \ket{m}
      \label{eq:PiOPE}\\
    & \equiv & i \pslash \gamma_{5} \left[ {\cal C}_{4} \ln(-p^{2}) +
        {\cal C}_{6} {1 \over p^{2}} + {\cal C}_{8} {1 \over p^{4}} +
\cdots
        \right] + i\gamma_{5} \left[ {\cal C}_{3} p^{2} \ln(-p^{2}) +
\cdots
        \right], \nonumber
\end{eqnarray}
where  $\bar{J}_N (0; t \sim 0.8)$ is for the $N^*$ state,
while $J_N(x; s = -1)$ for the $N$ state.
The correlation function (\ref{eq:PiOPE}) has been calculated up to
dimension 8, ignoring higher order terms in $m_{q}$ and $\alpha_{s}$.
The results are
\begin{eqnarray}
{\cal C}_{4} & \sim &
        m_{q} \bra{0} \bar{q}i \gamma_{5} q \ket{m}
        \stackrel{m_q \to 0}{\longrightarrow} 0 \, , \\
{\cal C}_{6} & = & -{s-t \over 4} \left[
    \langle \bar{d}d \rangle \bra{0} \bar{u} i \gamma_{5} u \ket{m} +
    \langle \bar{u}u \rangle \bra{0} \bar{d} i \gamma_{5} d \ket{m}
     \right] \, , \\
{\cal C}_{8} & = & - {s-t \over 144} \left[
    25 (\langle \bar{d}gG \cdot \sigma d \rangle
        \bra{0} \bar{u} i \gamma_{5} u \ket{m} +
    \langle \bar{u}gG \cdot \sigma u \rangle
        \bra{0} \bar{d} i \gamma_{5} d \ket{m}) \right.  \nonumber \\
    & &  \left. - 7 ( \langle \bar{d}d \rangle
        \bra{0} \bar{u} i \gamma_{5} gG \cdot \sigma u \ket{m} +
     \langle \bar{u}u \rangle
        \bra{0} \bar{d} i \gamma_{5} gG \cdot \sigma d \ket{m}) \right]
\, ,
\end{eqnarray}
where
$\langle \bar q q \rangle = \bra{0} \bar q q \ket{0}$
and
we have assumed 
the vacuum saturation for four-quark matrix elements.
To proceed further, we use the soft meson theorem:
\begin{equation}
    \bra{0} {\cal O}(0) \ket{m^{\alpha}(q)} \stackrel{q \rightarrow 0}
    {\longrightarrow} - {1 \over \sqrt{2} f_{m}} \int d^{3} x  \,
\bra{0}
    [i J^{\alpha}_{50}(x),{\cal O}(0)] \ket{0},
\end{equation}
where $J^{i}_{5\mu}(x)
= \bar{q}(x) \gamma_{\mu} \gamma_{5} (\lambda^{i} / 2) q(x)$
and $f_m$ is the decay constant of the meson $m$.
Thus we obtain the following relations for the operators 
$X = 1$ or $G\sigma$:
\begin{eqnarray}
  \label{eq:uum}
    \bra{0} \bar{u}i \gamma_{5} X u \ket{m} & = & -
        {\alpha_{m} \over f_{m}} \langle \bar{u} X u \rangle \, , \\
  \label{eq:ddm}
    \bra{0} \bar{d}i \gamma_{5} X d \ket{m} & = & \pm
        {\alpha_{m} \over f_{m}}  \langle \bar{d} X d  \rangle \, , 
\end{eqnarray}
where $\alpha_{\pi} = 1/\sqrt{2}$ and $\alpha_{\eta} = 1/\sqrt{6}$.
In (\ref{eq:ddm}), the plus sign is for the pion and the minus sign 
for the eta,  
because of the different isospin structure of the wave functions:  
$\pi^{0} \sim \frac{1}{\sqrt{2}}(\bar{u}u - \bar{d}d)$,
while $\eta \sim \eta_8
\sim \frac{1}{\sqrt{6}}(\bar{u}u + \bar{d}d - 2 \bar{s}s)$.  
For the eta matrix element, we have neglected a small mixing effect.
We note that the $\bar s s$ component in $\eta$ is irrelevant when 
$\alpha_S$ corrections are ignored, since the
interpolating field (\ref{eq:nucur}) does not contain
strange quarks.
From (\ref{eq:PiOPE}) -- (\ref{eq:ddm}), 
we find that the correlation function for the $\pi NN^{*}$ coupling
vanishes identically, while
that for the $\eta NN^*$ coupling remains finite.

Vanishing of the $\pslash \gamma_5$ term in the 
correlation function for $\pi NN^*$ is, in 
fact, a general consequence of chiral symmetry.  
We might have applied the soft meson theorem to the correlation function 
(\ref{eq:cor}) from the beginning.  
Using the transformation property 
$[ Q_5^a , J_N] = i \gamma_5 \tau^a J_N$, we find 
\begin{eqnarray}
\Pi^{\pi^a} (p) &=& \lim_{q \to 0}  
\int d^4x e^{ipx} \bra{0} T J_N(x) \bar J_N(0) \ket{\pi^a(q)} 
\nonumber \\
&=& - \frac{i}{\sqrt{2}f_\pi} \int d^4x e^{ipx} \bra{0} [ Q_5^a , 
T J_N(x) \bar J_N(0) ] \ket{0} \nonumber \\
&=&  \frac{1}{\sqrt{2}f_\pi} 
\int d^4x e^{ipx} 
\{ \gamma_5 , \bra{0} T J_N(x) \bar J_N(0)  \ket{0} \} \tau^a \, .
\label{Q5comm}
\end{eqnarray}
In the last expression, we have recovered the vacuum to vacuum
transition, 
which has a Lorentz structure 
$\bra{0}  J_N(x) \bar J_N(0)  \ket{0} \sim A \pslash + B 1$.  
Thus in (\ref{Q5comm}), the term of $\pslash \gamma_5$ disappears. 
We emphasize that the vanishing $\pslash \gamma_5$ term 
is a consequence of chiral symmetry.  
In order to achieve this result, the nucleon current has to
transform appropriately under chiral transformations and also, the
negative 
parity baryons are assumed to be produced by the same nucleon current 
$J_N$.
These two requirements determine chiral properties of the positive and 
negative parity nucleons.  
As a consequence, the $\pslash \gamma_5$ term, which in 
general should exist in the two point function, is shown to disappear.

What are then implications of the vanishing matrix element of 
$\pslash \gamma_5$ term?  
By relating this with the sum rule, it is implied that the coherent 
sum over all coupling constants of the pion between various baryon 
excitations will vanish. 
One would, in fact, 
make a stronger statement by using properties of analytic functions.  
Namely, if an analytic function vanishes in a certain domain, it
vanishes 
identically for whole analytic region.  
In the spirit of the QCD sum rule, we have found that the 
$\pslash \gamma_5$ term of the correlation 
function vanishes in the deep Euclidean region, which implies the  
identically vanishing $\pslash \gamma_5$ term for whole momentum space.  
There might be an exceptional case, if there are two 
(delta-function like) terms having equal strengths but
with opposite signs, which sum up to be zero.  
In fact, the low-lying negative parity nucleons do look like that; 
there are two neighboring states of $N(1535)$ and $N(1650)$.  
Even in this case, one would say that
a state which is coupled by the current $J_N$ is a 
superposition of two (or more) degenerate states, 
and couplings with the pion is 
coherently added up and cancel.  

For the eta case, we do not find a relation similar to (\ref{Q5comm}), 
because the current $J_N$ is not a good eigen state of the U(1) axial 
charge.  
Thus we find a finite contribution to the correlation function, which in 
turn is used in the sum rule analysis to extract a coupling constant 
$\geNR$. 
We have performed such a sum rule analysis, but the result turned 
out to be strongly parameter dependent.  
We have found that the coupling constant varies within 
$5 \lsim \geNR \lsim 20$, which is too large as compared with 
experimental value $\geNR \sim 2$.

Now, let us compare the above results with those of 
low energy effective models.  
We briefly discuss the nonrelativistic quark model, the large-$N_c$
method and an effective chiral lagrangian approach.

In the nonrelativistic quark model, the negative parity nucleon is
formed
by putting one of the valence quarks in the $p$ ($l = 1$)
orbit~\cite{ik}.
Therefore, there are two independent states for $J^P = 1/2^-$:
$\ket{1} = [l=1, S = 1/2]^{J=1/2}$ and
$\ket{2} = [l=1, S = 3/2]^{J=1/2}$,
where $S$ is the total intrinsic spin of the three quarks.
The physical state for $N(1535)$ is a linear combination of
these two states.
The coupling constants are the matrix elements of the operators,
${ \cal O}_\pi^\alpha = \sum_{i=1}^{3} \vec \sigma (i) \cdot
\vec \nabla (i) \tau^\alpha$ for the $\pi NN^*$  and
${ \cal O}_\eta = \sum_{i=1}^{3} \vec \sigma (i) \cdot
\vec \nabla (i) $ for the $\eta NN^*$.
The relative phase of the two states
$\ket{1}$ and $\ket{2}$ are then determined by the sign
of the tensor force.
In the Isgur-Karl model, it is given by the one gluon exchange
potential, while in a more sophisticated model, there is
a significant contribution from the one pion exchange
potential also~\cite{asy}.
In both cases the phase is given such that
the interference acts destructively for $\pi NN^*$ while
constructively for $\eta NN^*$.
This explains the relatively suppressed $\gpNR$.

The suppression of $\gpNR$ is also observed
in the large-$N_c$ limit.
Assuming that the lowest baryon state develops the hedgehog
intrinsic state with $K = J + I = 0$ (here the hedgehog has negative
parity as $J^P = 1/2^-$),
it is possible to show that the matrix element for
the $\gpNR$ coupling
$\bra{N} { \cal O}_\pi^a \ket{N^*}$ is of higher order in
$1/N_c$ as compared with that of $\bra{N} { \cal O}_\eta \ket{N^*}$.
An interesting observation here is that the negative parity hedgehog of 
$K=0$, $\ket{H^-}$, is simply given by 
$\ket{H^-} = \vec \sigma \cdot \hat r \ket{H^+}$.  
This is nothing but the non-relativistic reduction of 
$\gamma_5 \ket{H^+}$, suggesting that the negative parity 
hedgehog $\ket{H^-}$ would be a chiral partner of the ground state
hedgehog 
$\ket{H^+}$ in the large-$N_c$ limit.  
Furthermore, it is also possible to decompose the nucleon state
projected 
out from $\ket{H^-}$ and show that the wave function 
has the structure $\sim \ket{1} - \ket{2}$ when $N_c = 3$, where 
$\ket{1}$ and $\ket{2}$ are the two independent states of the 
non-relativistic quark model.  
Here, there is a strong similarity between 
the non-relativistic quark model and the large-$N_c$ method.  

Again turning to the matrix element for $\gpNR$ and $\geNR$, we see that 
\begin{eqnarray}
\gpNR &\sim&
\bra{N} \vec \sigma \cdot \hat r \;  \tau^a \ket{N^*} 
\sim \bra{N} \vec \sigma \cdot \hat r \;  
           \tau^a \vec \sigma \cdot \hat r \ket{N}
     \sim \bra{N} \tau^a \ket{N} \, , \nonumber \\
\geNR &\sim&
\bra{N} \vec \sigma \cdot \hat r  \ket{N^*} 
\sim \bra{N} \vec \sigma \cdot \hat r \;  
           \vec \sigma \cdot \hat r \ket{N}
\sim \bra{N} 1 \ket{N^*} \, .
\end{eqnarray}
In $\gpNR$, the operator $\tau^a$ in the last expression has the rank 
$I = 1$, $J = 0$, 
while in the second equation $I = J = 0$.  
The suppression of the $\gpNR$ coupling in the $1/N_c$ expansion is then
a 
consequence of the familiar $I=J$ rule for
large-$N_c$ baryons~\cite{largeN}.

One may wonder if such a suppression of $\gpNR$
could be explained in some way by spontaneously broken chiral symmetry.  
From chiral symmetry point of view, it seems natural to put $N$ and
$N^*$ 
in the same multiplet of chiral partner (or parity doublet).  
There have been several attempts to treat positive and negative
parity baryons in this point of view.  

DeTar and Kunihiro considered the parity doublet nucleons in the linear
sigma model of $SU(2) \times SU(2)$~\cite{DK}.
In addition to the standard chiral invariant interaction terms,
they introduced a chiral invariant mass term
between the positive and negative parity baryons.
The strength $m_{0}$ for the non-standard mass term
reduces to the mass of the would-be chiral doublet nucleons when the
chiral symmetry restores.
In the spontaneously broken phase,
the mass splitting is proportional to the non-zero 
value of the sigma condensate. 
In this model, it has been shown that
$\gpNR$ is proportional to $m_{0}$  to the leading order
in $m_{0}$.
Therefore, if $m_0=0$, the coupling constant $\gpNR$ vanishes.  
This is a rigorous consequence from chiral symetry.  
The question, is therefore, whether the non-standard mass term exists or 
not in the real world.  
This can be examined by looking at baryon masses in the chiral symmetric 
phase.  
In the QCD sum rule study~\cite{jko}, the masses of $N$
and $N^*$ get degenerate in the chiral symmetric phase with very small 
(possibly zero) masses.
This implies a small (and possibly vanishing) $m_{0}\approx 0$ 
and so a small  
$\gpNR$.

The formulation of DeTar and Kunihiro can be extended to the chiral U(1) 
$\times$ SU(2) model, where the eta is introduced as a chiral field of
the 
U(1) group.  
In this extension, it is possible to reproduce
the relation among the coupling constants:
$\gpNN >> \geNN$, while $\gpNR << \geNR$.

In summary, we have studied the $\pi NN^*$ and $\eta NN^*$ coupling
constant in the QCD sum rule.
The two point correlation function has been calculated in the soft
meson limit and up to dimension eight in the OPE.
To this order
$\gpNR $ is of order of $ {\cal O}(m_{q}, \alpha_{S})$ and therefore
is strongly suppressed.
The results in the QCD sum rule
are consistent with the predictions of the low energy effective
models such as the nonrelativistic quark model 
and of the large-$N_c$ method.
More interestingly, the suppression of
$\gpNR$ is also supported by spontaneously broken chiral 
symmetry by assuming the presence of the parity doublet baryons.

\end{document}